%% file: main.tex
\documentclass[10pt,twocolumn,letterpaper]{article}

\usepackage[pagenumbers]{cvpr}

\input{preamble}

\definecolor{cvprblue}{rgb}{0.21,0.49,0.74}
\usepackage[breaklinks,colorlinks,allcolors=cvprblue]{hyperref}

\def\paperID{arXiv}
\def\confName{CVPR}
\def\confYear{2026}

\title{How Many Visual Levers Drive Urban Perception?\\
Interventional Counterfactuals via Multiple Localised Edits}

\author{
Jason Tang \quad Stephen Law\\
University College London (UCL)\\
{\tt\small jasoncpits@outlook.com, stephen.law@ucl.ac.uk}
}

\begin{document}
\maketitle
\input{sec/0_abstract}    
\input{sec/1_intro}
\input{sec/2_related_work}
\input{sec/3_method}
\input{sec/4_results}
\input{sec/6_discussion}
\input{sec/7_conclusion}

{
    \small
    \bibliographystyle{ieeenat_fullname}
    \bibliography{main}
}
\input{sec/8_appendix}

\end{document}

%% file: preamble.tex
\usepackage{float}
\usepackage[most]{tcolorbox}
\newtcolorbox{promptbox}[1]{
  colback=black!5, colframe=black!30, boxrule=0.4pt,
  fonttitle=\small\bfseries, title={#1},
  left=4pt, right=4pt, top=2pt, bottom=2pt,
  before skip=6pt, after skip=6pt}



%% file: sec/0_abstract.tex
\begin{abstract}
    Street-view perception models predict subjective attributes such as safety at scale, but remain correlational: they do not identify which localized visual changes would plausibly shift human judgement for a specific scene. We propose a lever-based interventional counterfactual framework that recasts scene-level explainability as a bounded search over structured counterfactual edits. Each lever specifies a semantic concept, spatial support, intervention direction, and constrained edit template. Candidate edits are generated through prompt-conditioned image editing and retained only if they satisfy validity checks for same-place preservation, locality, realism, and plausibility. In a pilot across 50 scenes from five cities, the framework reveals preliminary proxy-based directional patterns and a practical failure taxonomy under prompt-only editing, with Mobility Infrastructure and Physical Maintenance showing the largest auxiliary safety shifts. Human pairwise judgements remain the ground-truth endpoint for future validation.
\end{abstract}

%% file: sec/1_intro.tex
\section{Introduction}
\label{sec:intro}

Street-view perception models predict subjective attributes such as safety, wealth, and liveliness at scale~\cite{salesses2013,naik2014,dubey2016}, but they remain limited as explanatory tools: they do not show which local visual changes would plausibly shift judgement for a specific scene. Existing explainability methods, including saliency, SHAP, and concept-based probes, are largely correlational, identifying features associated with a score without testing whether manipulating them changes perceived safety~\cite{adebayo2018,kim2018}. Urban-design research links cues such as greenery, maintenance, and visibility to perceived safety\footnote{\emph{Perceived safety} here refers to the subjective feeling of safety from crime and disorder, as elicited in Place Pulse~\cite{salesses2013} and SPECS~\cite{quintana2025}, rather than traffic safety. Some lever families, especially Mobility Infrastructure, may also evoke traffic-safety cues; whether these generalise to crime-safety perception remains open.}~\cite{li2015greenery,jiang2018brokenwindows,portnov2020lighting}, but does not establish which scene-specific local changes validly shift perception.

Following urban planning theory~\cite{jacobs1961,newman1972}, we study a restricted question: whether a scene admits multiple distinct \emph{single-lever} interventions that, evaluated independently against the same original image, plausibly shift perception. We propose an \emph{interventional counterfactual} protocol that makes this question testable through structured lever interventions defined by a semantic concept, spatial support, intervention direction, and constrained edit template. Candidate edits are generated with prompt-conditioned diffusion editing and retained only if a vision--language critic judges them to preserve the same place, remain localised, and appear realistic and plausible. Because generative editors may introduce non-target changes, each edit is treated as an auditable hypothesis rather than faithful evidence by default.

This paper introduces an auditable pilot framework rather than a validated causal estimate of human perception. We ask: (i) which lever families show positive auxiliary shifts after validity auditing, and (ii) how many distinct single-lever edits remain promising per scene after screening and thresholding?

Our contributions are:
\begin{enumerate}
    \item a \textbf{lever-based interventional counterfactual} formulation for street-view perception, with structured lever interventions $l=(c,s,d,\tau)$ as the explanatory unit;
    \item a \textbf{scene-specific generation-and-audit pipeline} that instantiates, realises, and validity-checks prompt-only edits; and
    \item a \textbf{50-scene pilot benchmark} across five cities, reporting preliminary proxy-based directional patterns and feasibility diagnostics under prompt-only editing.
\end{enumerate}

%% file: sec/2_related_work.tex
\section{Related Work}

\noindent\textbf{Urban perception and design cues.}
Crowdsourced pairwise judgement datasets have enabled vision models for
perceived safety, wealth, and liveliness at
scale~\cite{salesses2013,naik2014,dubey2016,quintana2025}, but these
models are not designed to answer mechanistic scene-level questions.
Urban-design theory links perceived safety to surveillance, active fontages,
upkeep, and legibility~\cite{jacobs1961,newman1972,loewen1993}, and
empirical street-view studies confirm roles for greenery, street-facing
windows, maintenance, and lighting~\cite{li2015greenery,denadai2016lively,jiang2018brokenwindows,portnov2020lighting}. However, none provides scene-level tests of localised, validity-audited
single-lever interventions.

\noindent\textbf{Interpretability and counterfactual explanations.}
Saliency and SHAP-style methods are known to be visually plausible yet
unfaithful~\cite{adebayo2018}; concept-based methods such as
TCAV~\cite{kim2018} remain correlational unless paired with explicit
interventions. Counterfactual visual explanations~\cite{goyal2019} and
causal evaluation benchmarks~\cite{melistas2024} provide stronger tools
but typically target model logits, focusing on the effect of a single feature on the model output rather than the holistic perceptual experience of the scene which can be affected by multiple single visual changes.

\noindent\textbf{Urban counterfactuals and diffusion-based editing.}
Law et al.~\cite{law2023} demonstrate plausible counterfactuals for
urban image regressors, establishing that generative edits can serve as
explanations in urban analytics but does not explicitly identify which edits are plausible for a given scene. UrbanPhysicalDisorder-4K~\cite{vera2025} brings counterfactual reasoning
into urban safety via annotated disorder features, but operates in
feature space rather than through localised image edits with explicit
validity auditing. Zhao et al.~\cite{zhao2026} propose
perception-guided street-view generation, but target scene optimisation
rather than scene-level interventional counterfactual analysis.
Diffusion-based editors such as SDEdit~\cite{meng2022},
InstructPix2Pix~\cite{brooks2023}, and
Prompt-to-Prompt~\cite{hertz2023prompt} make localised edits
increasingly feasible, but can introduce non-target drift that demands
explicit validity auditing.

\noindent\textbf{Positioning.}
The diffusion-editing literature above optimises for editing
fidelity, how faithfully a model executes an instruction, whereas our work sits at this intersection: rather than attributing a scene-level judgement to pixels or concepts alone, we search over structured, editor-feasible single-lever edits and retain only those that pass explicit validity auditing.

%% file: sec/3_method.tex
\section{Method}
\label{sec:method}

\subsection{Problem Setup}

Let $x \in \mathcal{X}$ denote a street-view image and
$a \in \mathcal{A}$ a perceptual attribute (e.g.\ safety). We take the
explanatory atom to be a \emph{lever intervention}
\begin{equation}
  l = (c, s, d, \tau),
\end{equation}
where $c$ is a lever concept (e.g.\ greenery, graffiti), $s$ is the
grounded scene support, $d$ the intervention direction (add, remove,
repair), and $\tau$ a constrained edit template.

For image $x$, let $L(x)$ denote the scene-specific candidate set
instantiated from an editor-feasible intervention vocabulary. Our method identifies the
subset that can be realised as valid edits:
\begin{equation}
  V(x) = \{ l_i \in L(x) \mid \exists\, \tilde{x}_i \text{ passing the validity audit} \}.
\end{equation}
Each lever is tested independently against the unedited original; edits
are never composed sequentially.

\subsection{Phase 1: Lever Candidate Construction}

We define an \emph{editor-feasible intervention vocabulary}: a bounded
set of lever concepts chosen to satisfy three criteria:
\textbf{(i)}~\emph{theoretical grounding} in the urban perception
literature~\cite{jacobs1961,newman1972,li2015greenery,jiang2018brokenwindows,portnov2020lighting},
\textbf{(ii)}~\emph{semantic distinctness} so concepts remain
interpretable, and \textbf{(iii)}~\emph{prompt-only editability}
without requiring segmentation masks or major structural changes.
The full concept list is given in Appendix~\ref{sec:intervention_vocabulary}.

For each image, a VLM planner grounds applicable vocabulary concepts to
visible scene regions, producing scene-specific candidates such as
\emph{remove graffiti from this wall} or \emph{add modest greenery near
this entrance}. These grounded candidates form $L(x)$.

\subsection{Phase 2: Bounded Stochastic Intervention}

For each candidate $l_i$, a prompt-only image generator $G$ produces
\begin{equation}
  x'_{i,j} = G(x;\, c_i, s_i, d_i, \tau_i, \varphi_j),
\end{equation}
where $\varphi_j$ indexes stochastic draws under a bounded retry budget
$T$, with $G$ being a prompt-only image editing model.

Each candidate is subjected to an automated validity audit performed by
a vision--language model,
evaluating: \textbf{(1)}~same-place preservation,
\textbf{(2)}~locality to the intended support, \textbf{(3)}~realism,
and \textbf{(4)}~plausibility of the intended lever. If at least one
draw passes within $T$ attempts, the first accepted edit $\tilde{x}_i$
is retained and $l_i$ enters $V(x)$. Only edits passing all four
criteria enter the retained set, so auxiliary shifts are measured only
on edits with confirmed semantic and spatial integrity.
We report valid-edit coverage
$\mathrm{Coverage}(x) = |V(x)| / |L(x)|$ as a property of edit
realisability under bounded prompt control rather than an urban
perception signal.

\subsection{Phase 3: Model-based Evaluation}

For each accepted edit $\tilde{x}_i$, we compute
\begin{equation}
  \Delta_i = f_a(\tilde{x}_i) - f_a(x),
\end{equation}
where $f_a$ is a ViT-B/16 perception model pretrained on the MIT Place
Pulse~2.0 pairwise safety dataset~\cite{hou2024globalstreetscapes},
which outputs a continuous score on a 0--10 scale (higher = safer).
This is strictly a model-based evaluation, not the main estimand; $f_a$ has
not been tuned on edited images, so $\Delta_i$ should be treated as
a ranking signal. The proxy-shortlisted set is
\begin{equation}
  E_{\mathrm{aux}}(x,a) = \{ l_i \in V(x) \mid \Delta_i >
  \theta_{\mathrm{aux}} \},
\end{equation}
where $\theta_{\mathrm{aux}}$ is an exploratory threshold.
Human-grounded pairwise evaluation remains future
work.%
\footnote{Our method is constrained interventional search, not
pixel-level attribution or causal identification. ``Multiple levers''
means distinct interventions that each pass validity checks
independently; it does not imply additive effects. Locality is audited,
not pixel-guaranteed.}

%% file: sec/4_results.tex
\section{Preliminary Results}
\label{sec:results}

\noindent\textbf{Dataset and setup.}
We use SPECS (Street Perception Evaluation Considering Socioeconomics)~\cite{quintana2025}, a demographically balanced survey in which 1{,}000 participants from five countries rated street-view scenes on ten perceptual indicators. We sample $N{=}50$ scenes from five SPECS cities (Amsterdam, Abuja, San Francisco, Santiago, Singapore; 10 per city), stratified by baseline safety and visual complexity to cover a broad range of starting conditions. Each scene allows up to $K{=}5$ lever candidates with a stochastic budget of $T{=}5$ generation attempts per candidate. We instantiate the framework \footnote{The framework is model-agnostic: any component (e.g., planner, editor, auditor) can be swapped, finetuned, or extended as needed.} with Qwen-Image-Edit as the prompt-only editor and GPT-5.4 as the LLM-as-judge critic, yielding a reproducible baseline. Validity auditing acts as an eligibility gate: all reported effect summaries are computed only on edits retained in $V(x)$.

\noindent\textbf{Proxy safety-score shifts.}
For each valid edit $\tilde{x}_i$, we compute the model-based score delta $\Delta_i$ using the ViT-B/16 safety proxy defined in \S\ref{sec:method}; positive values indicate a higher
predicted safety score relative to the unedited original. Among the 177 valid edits, the proxy produces a mean shift of $+0.366$ (95\% CI $[+0.199, +0.537]$, median $+0.184$, range
$[-3.624, +3.842]$). Using threshold $\theta_{\mathrm{aux}}{=}0.1$,
95 of 177 valid edits fall in $E_{\mathrm{aux}}(x,a)$, corresponding to
a mean of 1.90 proxy-shortlisted levers per scene
(95\% CI $[1.476, 2.324]$); 40 of the 50 scenes retain at least one
proxy-shortlisted lever and 24 retain multiple.
The largest positive proxy shifts are observed for
lane-marking repainting, crosswalk repainting, and localized greenery
addition.

\begin{table}[t]
  \centering
  \caption{Lever-type results grouped by intervention family.
  Mean~$\Delta_{\mathrm{aux}}$ and $\Delta_{\mathrm{aux}}$ [95\% CI]
  are computed over valid edits only where $\theta_{\mathrm{aux}}{>=}0.1$; confidence intervals are
  bootstrap percentile intervals over the retained set.}
  \label{tab:levers}
  \scriptsize
  \resizebox{\columnwidth}{!}{%
  \begin{tabular}{p{0.43\linewidth}ccc}
    \toprule
    Family / Lever concept & Valid & Mean $\Delta_{\mathrm{aux}}$ & $\Delta_{\mathrm{aux}}$ [95\% CI] \\
    \midrule
    \textit{Physical Maintenance} & & & \\
    \quad Graffiti removal        & 4  & $+$0.396 & [$+$0.121, $+$0.670] \\
    \quad Litter removal          & 25 & $+$0.296 & [$-$0.128, $+$0.743] \\
    \quad Facade repair           & 4  & $+$0.519 & [$+$0.105, $+$0.949] \\
    \quad Surface cleaning        & 38 & $+$0.352 & [$+$0.026, $+$0.684] \\
    \quad\textbf{Family total}    & 71 & $+$0.344 & [$+$0.107, $+$0.584] \\
    \midrule
    \textit{Environmental Amenity} & & & \\
    \quad Localized greenery add.\  & 32 & $+$0.650 & [$+$0.241, $+$1.070] \\
    \quad Lighting repair           & 8  & $-$0.232 & [$-$0.598, $+$0.173] \\
    \quad Tree canopy management    & 14 & $-$0.245 & [$-$0.952, $+$0.446] \\
    \quad\textbf{Family total}      & 54 & $+$0.287 & [$-$0.041, $+$0.610] \\
    \midrule
    \textit{Visual Legibility} & & & \\
    \quad Signage decluttering         & 5 & $-$0.631 & [$-$1.294, $-$0.151] \\
    \quad Storefront transparency      & 2 & $+$0.958 & [$+$0.184, $+$1.732] \\
    \quad\textbf{Family total}         & 7 & $-$0.177 & [$-$0.927, $+$0.599] \\
    \midrule
    \textit{Mobility Infrastructure} & & & \\
    \quad Crosswalk repainting       & 15 & $+$0.767 & [$+$0.255, $+$1.261] \\
    \quad Lane marking repainting    & 30 & $+$0.485 & [$+$0.045, $+$0.934] \\
    \quad\textbf{Family total}       & 45 & $+$0.579 & [$+$0.244, $+$0.923] \\
    \midrule
    \textbf{Overall}                 & 177 & $+$0.366 & [$+$0.199, $+$0.537] \\
    \bottomrule
  \end{tabular}
  }
\end{table}

Mobility Infrastructure yields the largest family-level mean auxiliary
shift ($+0.579$), led by crosswalk and lane-marking repainting.
These interventions are conventionally associated with traffic safety
rather than crime safety; whether the proxy shift reflects a genuine
crime-safety signal or a scorer artefact remains an open question
(see safety-definition footnote in \S\ref{sec:intro}).
Physical Maintenance is the broadest consistently positive family
($+0.344$), driven by surface cleaning, litter removal, facade repair,
and graffiti removal. Environmental Amenity is more heterogeneous:
localized greenery addition is strongly positive ($+0.650$), whereas
lighting repair and tree canopy management are weak or
negative.\footnote{Tree canopy management refers to pruning or trimming
overgrown canopy to improve sightlines and pedestrian visibility, 
the opposite direction from greenery addition. The ViT-B/16 safety model, trained
on Place Pulse images where visible greenery correlates with higher
safety, may penalise any net reduction in canopy cover regardless of
the urbanistic intent.}
Visual Legibility is small-sample and mixed in sign; the negative mean
for signage decluttering ($-0.631$) may reflect the proxy associating
commercial signage density with activity in Place Pulse training data.
Per-lever proposal and valid-rate diagnostics are in
Appendix Tables~\ref{tab:family_appendix} and~\ref{tab:city_appendix}.

\noindent\textbf{Thresholded proxy-shortlisted lever counts.}
Appendix Figure~\ref{fig:delta_cutoff_counts} counts levers that are both valid
and directionally notable at increasingly strict auxiliary thresholds.
At $\theta_{\mathrm{aux}}{=}0.1$, Physical Maintenance averages 0.76
proxy-shortlisted levers per scene, Mobility Infrastructure 0.56, and
Environmental Amenity 0.54; Visual Legibility averages only 0.04.
By city, San Francisco and Santiago dominate the positive tail with
2.5 and 2.6 proxy-shortlisted levers per scene respectively, while
Abuja averages 1.1. As the cutoff rises to $1.0$, the family ranking
compresses but Environmental Amenity and Mobility Infrastructure retain
the largest positive tails.

\begin{table}[t]
  \centering
  \caption{City-level summary over the full $N{=}50$ run.
  Mean $\Delta_{\mathrm{aux}}$ and $\Delta_{\mathrm{aux}}$ [95\% CI]
  are computed over valid edits only. Coverage diagnostics are reported
  in Appendix Table~\ref{tab:city_appendix}.}
  \label{tab:cities}
  \scriptsize
  \resizebox{\columnwidth}{!}{%
  \begin{tabular}{lccc}
    \toprule
    City & Valid & Mean $\Delta_{\mathrm{aux}}$ & $\Delta_{\mathrm{aux}}$ [95\% CI] \\
    \midrule
    Amsterdam     & 40 & $+$0.400 & [$+$0.073, $+$0.764] \\
    Abuja         & 32 & $-$0.136 & [$-$0.441, $+$0.141] \\
    San Francisco & 36 & $+$0.704 & [$+$0.285, $+$1.105] \\
    Santiago      & 38 & $+$0.742 & [$+$0.398, $+$1.089] \\
    Singapore     & 31 & $-$0.012 & [$-$0.383, $+$0.352] \\
    \midrule
    \textbf{Overall} & 177 & $+$0.366 & [$+$0.199, $+$0.537] \\
    \bottomrule
  \end{tabular}
  }
\end{table}

Santiago and San Francisco show the largest positive mean shifts; Abuja
and Singapore are near zero or slightly negative. Valid-rate diagnostics
by city are in Appendix Table~\ref{tab:city_appendix}.

\noindent\textbf{Baseline score and gate-qualified editability.}
Appendix Figure~\ref{fig:baseline_vs_valid} plots each scene's baseline safety
score $f_a(x)$ against its valid lever count $|V(x)|$. At $N{=}50$, no
meaningful monotonic relationship is observed (Spearman
$\rho{=}0.10$, $p{=}0.511$): scenes with lower baseline safety do not
admit more valid levers than higher-baseline scenes. We revisit this
null pattern in \S\ref{sec:discussion}.

\noindent\textbf{Validity gate.}
Of 250 proposed candidates, 177 pass the validity audit (rate 0.708);
all 50 scenes retain at least one valid lever. Detailed coverage and
failure analysis are in Appendix~\ref{sec:validity_appendix};
representative rejected counterfactuals are shown in Appendix
Figure~\ref{fig:rejected_qualitative}.

\noindent\textbf{Qualitative examples.}
Three positive auxiliary-shift examples -- lane-marking repainting
in Amsterdam, facade repair in Singapore, and surface cleaning in
Santiago -- are shown in Appendix Figure~\ref{fig:qualitative}. All
three are spatially constrained, visually legible, and plausible as
single-lever urban changes.

%% file: sec/6_discussion.tex
\section{Discussion and Future Work}
\label{sec:discussion}

This pilot establishes two outputs: a feasibility profile for the generation-and-audit pipeline, and preliminary proxy-based directional patterns to prioritise for human evaluation. On feasibility, roadway-marking and maintenance interventions realise reliably across cities, while lighting repair remains the clearest bottleneck. No meaningful relationship is observed between baseline safety and valid lever count, suggesting that multi-lever richness depends more on scene geometry than starting safety level. On directionality, Mobility Infrastructure yields the largest retained auxiliary shifts, though it remains unclear whether this reflects human perceptual sensitivity, scorer priors, or edit realisability.

Environmental Amenity is more heterogeneous. Localised greenery addition is strongly positive, whereas lighting repair and canopy management are negative on average despite being theory-grounded cues. This contrast suggests that the auxiliary scorer may encode a simple “more green = safer” prior rather than a more nuanced sensitivity to sightlines or surveillance, and that family-level aggregation can obscure important lever-specific differences.

\noindent\textbf{On the model-based evaluation proxy.} The scorer $f_a$ has not been validated on edited images, so $\Delta_i$ should not be read as a calibrated effect size. However, for monotone interventions (road markings appear, graffiti is removed, greenery is added), the directional signal is grounded in large-scale crowdsourced pairwise preference over this class of scene variation, making the sign and relative magnitude of $\Delta_{\mathrm{aux}}$ a useful prioritisation signal for selecting which levers and scenes to send to human evaluation.

\noindent\textbf{XAI contribution.} Beyond urban perception, the methodological contribution is a stronger standard for visual explanation: the proposed change must be semantically coherent and generatively feasible before it enters the attribution result, providing a reusable scaffold for interventional XAI in other perceptual domains.

\noindent\textbf{Future directions and human endpoint.}
The present pipeline is intentionally minimal, prompt-only generation, a prompted VLM critic, and an auxiliary proxy scorer are used here to establish a reproducible baseline rather than a fully validated explanation system. Four directions follow naturally from the
pipeline structure. \emph{(i) Planner tuning:} learning to predict lever
viability from scene features, using the validity and
$\Delta_{\mathrm{aux}}$ outcomes reported here as supervision, so that
the $K$-candidate budget is spent on levers most likely to be both
realisable and promising under the auxiliary proxy. \emph{(ii) Generator alignment:} optimising
for validity-gate criteria via reward-from-feedback or mask-constrained
generation; the lighting-repair failure mode (valid rate 0.30) is the
clearest target, with per-family valid rates as the benchmark metric.
\emph{(iii) Critic calibration:} aligning the VLM critic with human
raters and ultimately learning a critic that predicts 2AFC preference
directly, reducing the gap between the auxiliary proxy and the human
endpoint. \emph{(iv) End-to-end optimisation:} jointly training planner,
generator, and critic once each component is independently validated.
The ground-truth estimand throughout is pairwise human safety ratings on passing edits, collected via a randomised 2AFC study. Until then,
validity should be read as an eligibility control on image generation
and auxiliary shifts as a prioritisation signal for which lever
families and cities to send to human evaluation first.

%% file: sec/7_conclusion.tex
\section{Conclusion}

The 50-scene pilot shows that structured, validity-audited single-lever edits can be generated at scale, with Mobility Infrastructure and Physical Maintenance showing the most consistent positive shifts and therefore the clearest priorities for human evaluation. More broadly, the framework shifts explanation from post-hoc pixel attribution to semantically grounded interventions. Realising that potential will require better planning, editing control, critic calibration, and ultimately pairwise human judgement as the ground-truth endpoint.

%% file: sec/8_appendix.tex
\newpage
\onecolumn
\appendix
\renewcommand{\thefigure}{A\arabic{figure}}
\renewcommand{\thetable}{A\arabic{table}}
\setcounter{figure}{0}
\setcounter{table}{0}

\section{Appendix}

\subsection{Pipeline and Intervention Vocabulary}
\label{sec:appendix_pipeline}

\begin{figure}[H]
  \centering
  \includegraphics[width=0.85\textwidth]{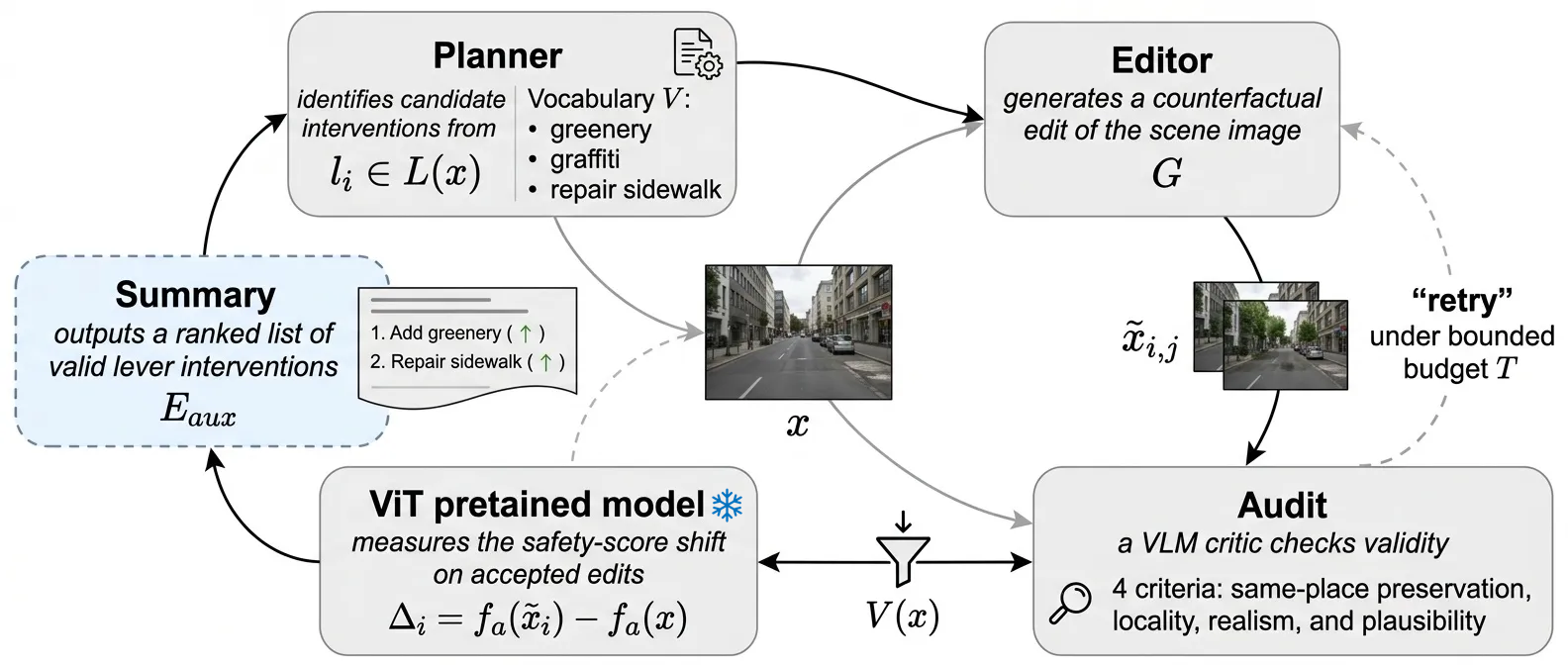}
  \caption{Overview of the lever-based interventional counterfactual
  pipeline.  Each candidate is a structured lever intervention
  $l=(c,s,d,\tau)$.  Phase~1 constructs scene-specific candidates from a
  curated ontology.  Phase~2 performs bounded stochastic generation
  subject to a four-criterion validity audit.  Phase~3 summarises
  accepted edits with auxiliary classifier scoring and, optionally,
  future human pairwise judgements.}
  \label{fig:pipeline}
\end{figure}

\begin{table}[H]
  \centering
  \caption{Intervention vocabulary.  Each concept satisfies all three
  inclusion criteria defined in \S\ref{sec:method}.  Structural
  interventions and social-order cues are excluded as they are not
  realisable under prompt-only editing.}
  \label{sec:intervention_vocabulary}%
  \small
  \begin{tabular}{@{} l l p{5.8cm} @{}}
    \toprule
    Family & Basis & Lever concepts \\
    \midrule
    Physical Maintenance
      & Broken-windows / upkeep~\cite{newman1972,jiang2018brokenwindows}
      & Graffiti removal; litter removal; facade repair; surface cleaning; shutter repair \\[4pt]
    Environmental Amenity
      & Biophilic safety / visibility~\cite{li2015greenery,portnov2020lighting}
      & Localised greenery addition; lighting repair; tree canopy management \\[4pt]
    Visual Legibility
      & Active-frontages / surveillance~\cite{jacobs1961}
      & Signage decluttering; storefront transparency increase \\[4pt]
    Mobility Infrastructure
      & Walkability / road markings~\cite{quintana2025}\textsuperscript{*}
      & Crosswalk repainting; lane marking repainting \\
    \bottomrule
    \multicolumn{3}{@{}l}{\footnotesize\textsuperscript{*}See safety-definition footnote in \S\ref{sec:intro}.}
  \end{tabular}
\end{table}

\subsection{Qualitative Examples}
\label{sec:appendix_qualitative}

\begin{figure}[H]
  \centering
  \includegraphics[width=\textwidth]{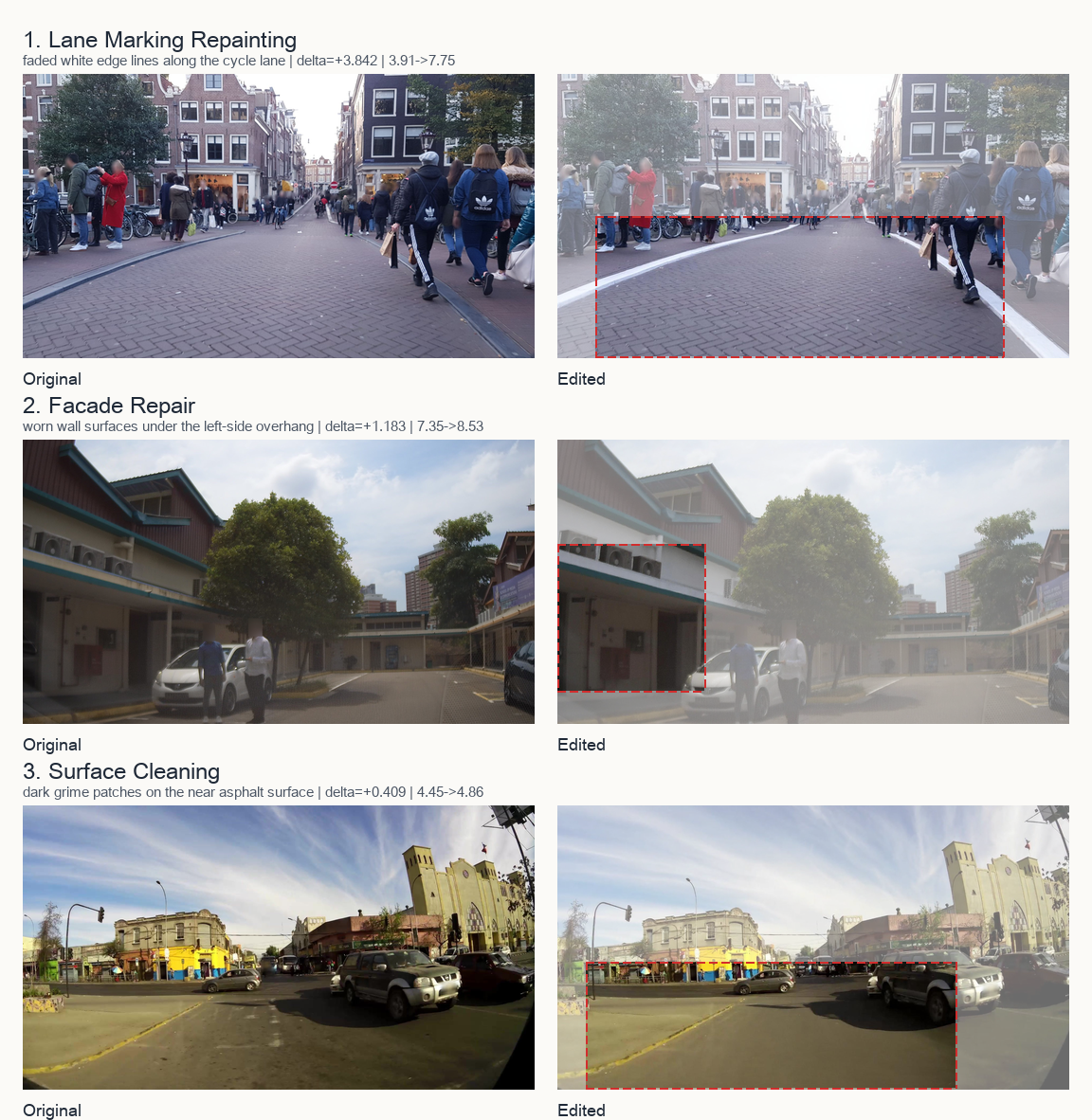}
  \caption{Three qualitative examples from the retained set.
  From top: lane-marking repainting (Amsterdam), facade repair
  (Singapore), surface cleaning (Santiago).  Red dashed bounding boxes
  indicate the edited region; surrounding areas are dimmed for contrast.}
  \label{fig:qualitative}
\end{figure}

\begin{figure}[H]
  \centering
  \includegraphics[width=\textwidth]{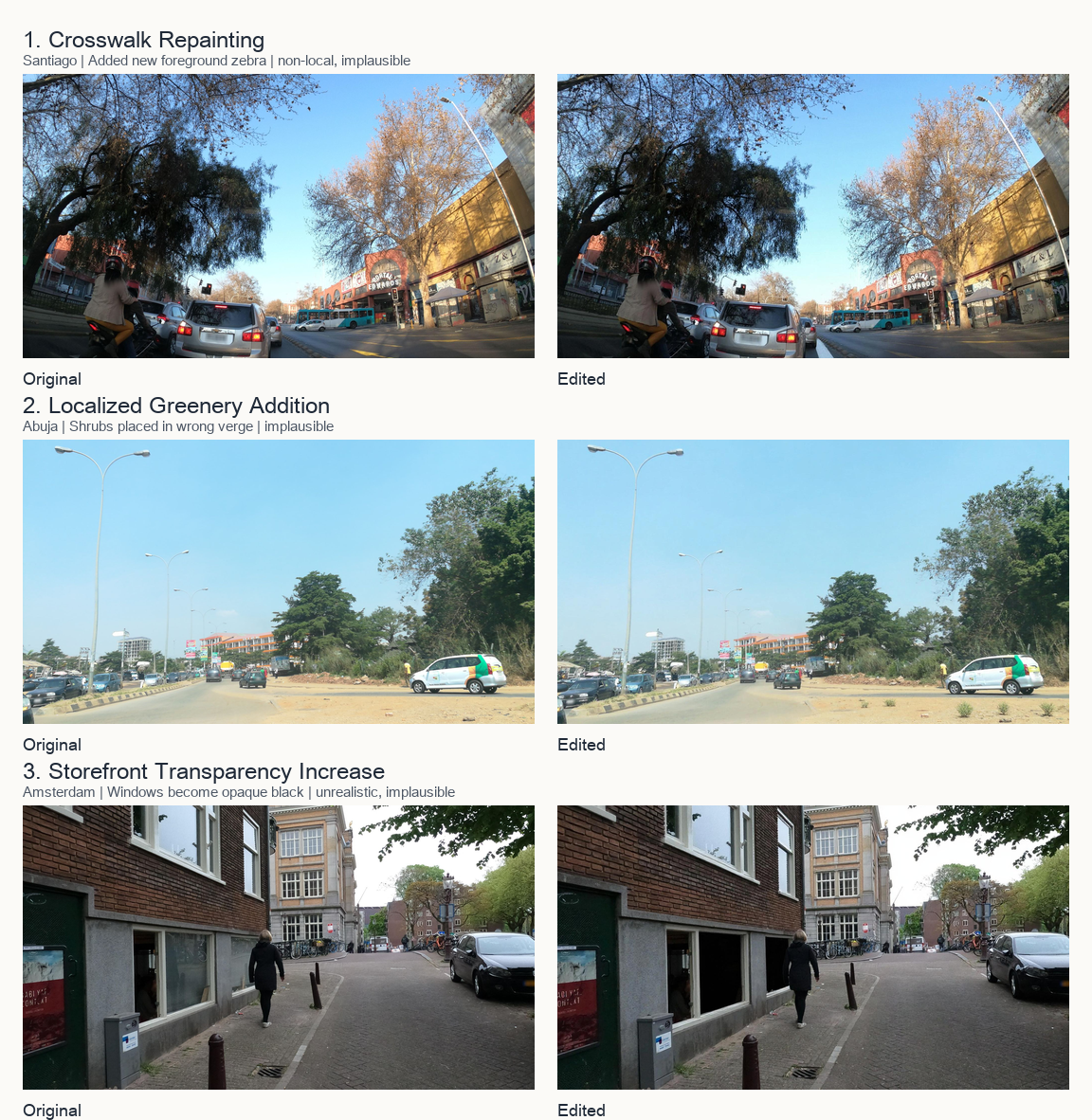}
  \caption{Three representative rejected counterfactuals from the
  audited-but-rejected set. From top: crosswalk repainting adds a new
  foreground zebra crossing rather than repairing the supported
  intersection marking (Santiago), localised greenery addition places
  shrubs in the wrong roadside verge (Abuja), and storefront
  transparency increase turns the target windows into opaque black
  openings rather than revealing a plausible interior (Amsterdam).}
  \label{fig:rejected_qualitative}
\end{figure}

\subsection{Validity and Coverage Details}
\label{sec:validity_appendix}

\begin{figure}[H]
  \centering
  \begin{minipage}[t]{0.48\textwidth}
    \centering
    \includegraphics[width=\linewidth]{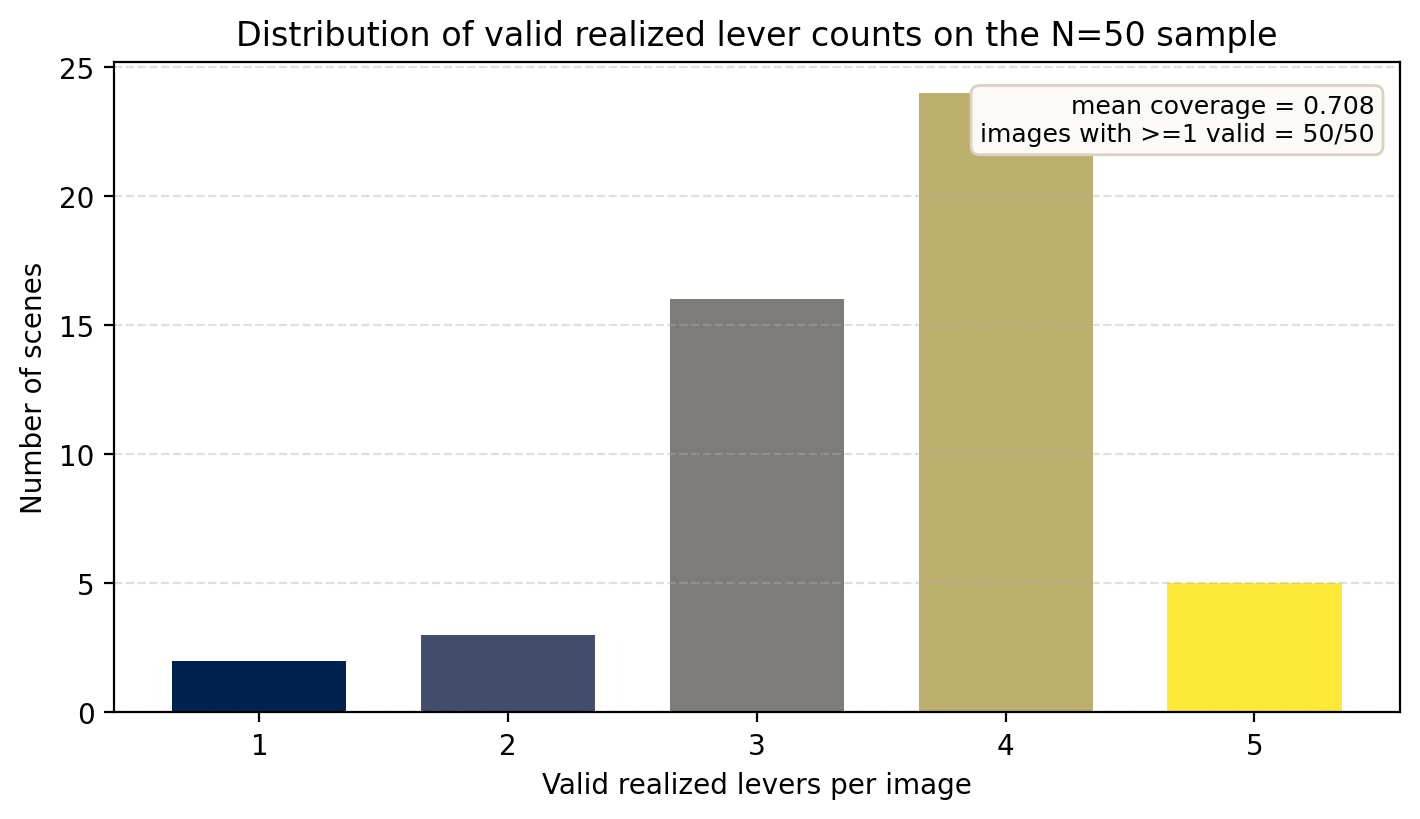}
    \captionof{figure}{Distribution of valid realised lever counts per
    scene ($N{=}50$).  Most scenes admit three or four independently
    valid single-lever interventions.}
    \label{fig:distribution}
  \end{minipage}\hfill
  \begin{minipage}[t]{0.48\textwidth}
    \centering
    \includegraphics[width=\linewidth]{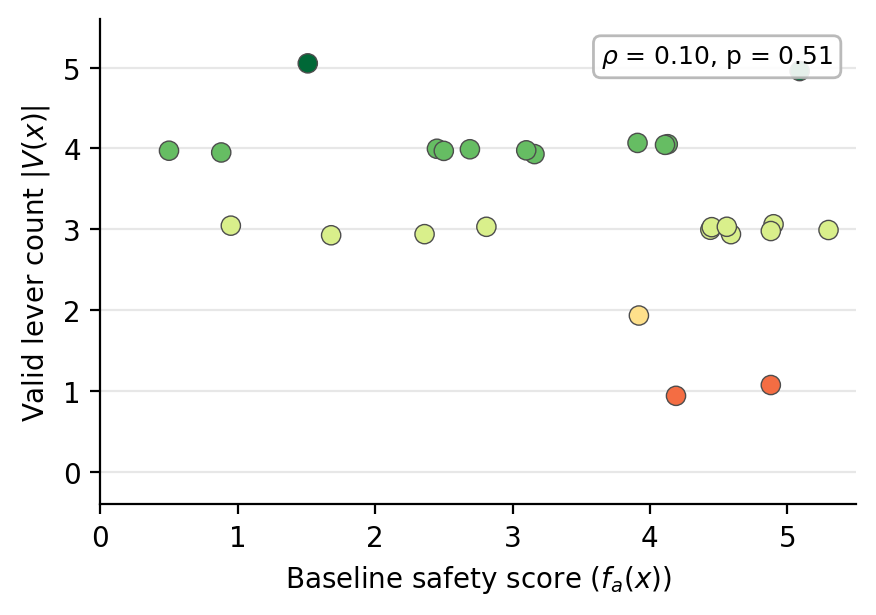}
    \captionof{figure}{Baseline safety score vs.\ valid lever count per
    scene ($N{=}50$).  No meaningful relationship is observed
    (Spearman $\rho{=}0.10$, $p{=}0.511$).}
    \label{fig:baseline_vs_valid}
  \end{minipage}
\end{figure}

Across 50 scenes the planner produces the full 250 candidate rows
(5.0 per image on average), of which 177 pass the validity audit,
yielding mean coverage $|V(x)|/|L(x)| = 0.708$.
Figure~\ref{fig:distribution} shows the distribution of valid counts per
scene: 2~scenes have one valid lever, 3~have two, 16~have three,
24~have four, and 5~have all five.
The dominant bottleneck is not candidate proposal but valid realisation.
Among the 73 audited-but-rejected edits, plausibility dominates
completely; 45 also show no discernible target change, 25 exhibit
non-local drift, 5 are unrealistic, and none fail same-place
preservation.  Representative rejected counterfactuals are shown in
Figure~\ref{fig:rejected_qualitative}. Generator failure before audit
is zero in the cleaned
$N{=}50$ run, so the remaining variance is concentrated in valid
realisation rather than missing candidate production.

\begin{table}[H]
  \centering
  \caption{Family-level summary over the full $N{=}50$ run.
  Valid-rate intervals are 95\,\% Wilson confidence intervals; mean
  $\Delta_{\mathrm{aux}}$ and its 95\,\% CI are
  computed over valid edits only.
  $\Delta_{\mathrm{aux}}$ values are on the 0--10 proxy scale.}
  \label{tab:family_appendix}
  \small
  \begin{tabular}{@{} l c c c c @{}}
    \toprule
    Family & Valid / Prop. & Rate [95\,\% CI] & Mean $\Delta_{\mathrm{aux}}$ & $\Delta_{\mathrm{aux}}$ [95\,\% CI] \\
    \midrule
    Physical Maintenance    & 71\,/\,92  & 0.77\,[0.68, 0.85] & $+$0.344 & [$+$0.107, $+$0.584] \\
    Environmental Amenity   & 54\,/\,82  & 0.66\,[0.55, 0.75] & $+$0.287 & [$-$0.041, $+$0.610] \\
    Visual Legibility       &  7\,/\,10  & 0.70\,[0.40, 0.89] & $-$0.177 & [$-$0.927, $+$0.599] \\
    Mobility Infrastructure & 45\,/\,66  & 0.68\,[0.56, 0.78] & $+$0.579 & [$+$0.244, $+$0.923] \\
    \midrule
    \textbf{Overall}        & 177\,/\,250 & 0.71\,[0.65, 0.76] & $+$0.366 & [$+$0.199, $+$0.537] \\
    \bottomrule
  \end{tabular}
\end{table}

\begin{table}[H]
  \centering
  \caption{City-level coverage and directional summary over the full
  $N{=}50$ run.  Each city contributes 10~scenes and therefore
  50~proposed candidates.  Valid-rate intervals are 95\,\% Wilson
  confidence intervals; mean $\Delta_{\mathrm{aux}}$ and its 95\,\% CI
  are computed over valid edits only.
  $\Delta_{\mathrm{aux}}$ values are on the 0--10 proxy scale.}
  \label{tab:city_appendix}
  \small
  \begin{tabular}{@{} l c c c c @{}}
    \toprule
    City & Valid / Prop. & Rate [95\,\% CI] & Mean $\Delta_{\mathrm{aux}}$ & $\Delta_{\mathrm{aux}}$ [95\,\% CI] \\
    \midrule
    Amsterdam     & 40\,/\,50  & 0.80\,[0.67, 0.89] & $+$0.400 & [$+$0.073, $+$0.764] \\
    Abuja         & 32\,/\,50  & 0.64\,[0.50, 0.76] & $-$0.136 & [$-$0.441, $+$0.141] \\
    San Francisco & 36\,/\,50  & 0.72\,[0.58, 0.83] & $+$0.704 & [$+$0.285, $+$1.105] \\
    Santiago      & 38\,/\,50  & 0.76\,[0.63, 0.86] & $+$0.742 & [$+$0.398, $+$1.089] \\
    Singapore     & 31\,/\,50  & 0.62\,[0.48, 0.74] & $-$0.012 & [$-$0.383, $+$0.352] \\
    \midrule
    \textbf{Overall} & 177\,/\,250 & 0.71\,[0.65, 0.76] & $+$0.366 & [$+$0.199, $+$0.537] \\
    \bottomrule
  \end{tabular}
\end{table}

\begin{figure}[H]
  \centering
  \includegraphics[width=0.85\textwidth]{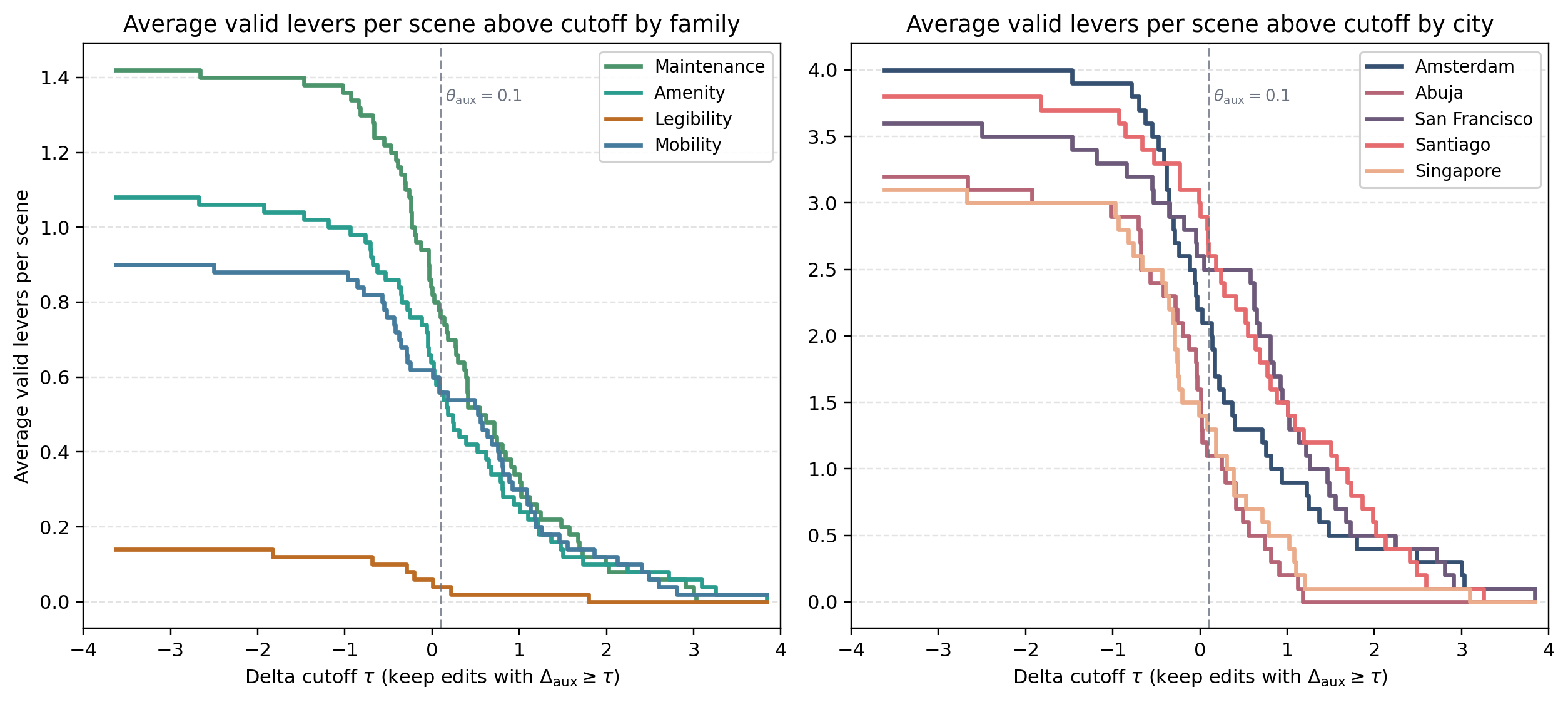}
  \caption{Average number of retained levers per scene satisfying
  $\Delta_{\mathrm{aux}} \geq \tau$, split by lever family (left) and
  city (right).  The vertical line marks the operating threshold
  $\theta_{\mathrm{aux}}{=}0.1$.  Annotations report the overall mean
  $\Delta_{\mathrm{aux}}$ with 95\,\% CI and the mean number of
  proxy-shortlisted levers per scene with 95\,\% CI.}
  \label{fig:delta_cutoff_counts}
\end{figure}

\subsection{Prompt Contracts}
\label{sec:prompt_contracts}

The planner, editor, and critic each operate under a structured prompt
contract.  Condensed versions are shown below; full prompts are in the
released repository.

\begin{promptbox}{Planner prompt (condensed)}
\scriptsize\ttfamily
You are an urban perception planner. Given a street-view image and
target percept, propose a constrained set of candidate lever
interventions.\par\medskip
ONTOLOGY: Choose only from four families --- Physical Maintenance,
Environmental Amenity, Visual Legibility, Mobility Infrastructure ---
and prefer cross-family diversity when the scene supports it.\par\medskip
HARD CONSTRAINTS: (1) One lever per candidate. (2) Grounded in a
visible scene element. (3) Local, plausible, prompt-only friendly.
(4) No global relighting, weather, or camera changes. (5) Prefer
the smallest plausible intervention. (6) Exclude theoretically
relevant levers whose target element is not clearly visible/editable
in the image.\par\medskip
DIVERSITY: Return distinct candidates using different lever concepts
and avoid magnitude variants of the same intervention.\par\medskip
Return JSON with field ``candidates'', each containing:
lever\_concept, lever\_family, scene\_support, target\_object,
intervention\_direction, edit\_template, edit\_plan.
\end{promptbox}

\begin{promptbox}{Edit prompt (condensed)}
\scriptsize\ttfamily
Use the PROVIDED image as base. Preserve exact viewpoint, geometry,
and layout.\par\medskip
ALLOWED: Only modify the target object as required by the plan.
If repainting/retexturing, keep shape and placement identical.\par\medskip
FORBIDDEN: No global restyling, relighting, or recoloring. No
adding/removing other objects. No readable text. No background,
sky, road, or context changes.\par\medskip
Lever concept: \{lever\_concept\} \\
Lever family: \{lever\_family\} \\
Scene support: \{scene\_support\} \\
Intervention direction: \{intervention\_direction\} \\
Edit template: \{edit\_template\} \\
Target object: \{target\_object\} \\
Edit plan: \{edit\_plan\}
\end{promptbox}

\begin{promptbox}{Critic prompt (condensed)}
\scriptsize\ttfamily
Evaluate whether the edited image is a valid single-lever
counterfactual relative to the original.\par\medskip
CONTEXT: Edit produced by prompt-only diffusion without masking.
Minor incidental changes (tone drift, texture resampling) are
expected artefacts and should not cause failure unless they
materially alter scene meaning; plausibility and locality are still
judged strictly against the requested lever and support.\par\medskip
CRITERIA:\par\smallskip
(A) edit\_attempted: generator made a visible change at the target
    (false only if output looks identical to original there).\par
(B) same\_place\_preserved: same underlying place.\par
(C) is\_localised: primary meaningful change is in/near intended
    support; minor global tone shifts do not count as non-local.\par
(D) is\_realistic: physically plausible and coherent.\par
(E) is\_plausible: recognisably the requested lever at the stated
    support; fail if the edit type/support is wrong or too
    excessive.\par\medskip
CLEAR FAIL CONDITIONS: Viewpoint/geometry changed; large non-target
objects added/removed; requested lever replaced by a different
change; no discernible change at target.\par\medskip
Return JSON: \{edit\_attempted, same\_place\_preserved,
is\_localised, is\_realistic, is\_plausible, notes\}, where
notes = \{failure\_modes, diagnosis, repair\_suggestion\}.
\end{promptbox}